\documentclass[sigconf]{acmart}
\AtBeginDocument{%
  \providecommand\BibTeX{{%
    \normalfont B\kern-0.5em{\scshape i\kern-0.25em b}\kern-0.8em\TeX}}}

\copyrightyear{2023}
\acmYear{2023}
\setcopyright{rightsretained}
\acmConference[FAccT '23]{2023 ACM Conference on Fairness, Accountability, and Transparency}{June 12--15, 2023}{Chicago, IL, USA}
\acmBooktitle{2023 ACM Conference on Fairness, Accountability, and Transparency (FAccT '23), June 12--15, 2023, Chicago, IL, USA}
\acmDOI{10.1145/3593013.3593990}
\acmISBN{979-8-4007-0192-4/23/06}

\usepackage{hyperref}
\usepackage{breakurl}

\begin{document}

%%
%% The "title" command has an optional parameter,
%% allowing the author to define a "short title" to be used in page headers.
\title [Walking the Walk of AI Ethics] {Walking the Walk of AI Ethics: Organizational Challenges and the Individualization of Risk among Ethics Entrepreneurs}

%%
%% The "author" command and its associated commands are used to define
%% the authors and their affiliations.
%% Of note is the shared affiliation of the first two authors, and the
%% "authornote" and "authornotemark" commands
%% used to denote shared contribution to the research.
\author{Sanna J. Ali}
\affiliation{%
  \institution{Stanford University}
  \city{Stanford}
  \state{CA}
  \country{USA}
  \postcode{94304}
  }
\email{sanna@alumni.stanford.edu}
\author{Angèle Christin}
\affiliation{%
  \institution{Stanford University}
  \city{Stanford}
  \state{CA}
  \country{USA}
  \postcode{94304}
  }
\email{angelec@stanford.edu}

\author{Andrew Smart}
\affiliation{%
  \institution{Google Research}
  \city{San Francisco}
  \state{CA}
  \country{USA}}
\email{andrewsmart@google.com}

\author{Riitta Katila}
\affiliation{%
  \institution{Stanford University}
  \city{Stanford}
  \state{CA}
  \country{USA}
  \postcode{94304}
  }
\email{rkatila@stanford.edu}

%%
%% By default, the full list of authors will be used in the page
%% headers. Often, this list is too long, and will overlap
%% other information printed in the page headers. This command allows
%% the author to define a more concise list
%% of authors' names for this purpose.
\renewcommand{\shortauthors}{Ali, Christin, Smart, Katila}

%%
%% The abstract is a short summary of the work to be presented in the
%% article.
\begin{abstract}
Amidst decline in public trust in technology, computing ethics have taken center stage, and critics have raised questions about corporate “ethics washing.” Yet few studies examine the actual implementation of AI ethics values in technology companies. Based on a qualitative analysis of technology workers tasked with integrating AI ethics into product development, we find that workers experience an environment where policies, practices, and outcomes are decoupled. We analyze AI ethics workers as \textit{ethics entrepreneurs} who work to institutionalize new ethics-related practices within organizations. We show that ethics entrepreneurs face three major barriers to their work. First, they struggle to have ethics prioritized in an environment centered around software product launches. Second, ethics are difficult to quantify in a context where company goals are incentivized by metrics. Third, the frequent reorganization of teams makes it difficult to access knowledge and maintain relationships central to their work. Consequently, individuals take on great personal risk when raising ethics issues, especially when they come from marginalized backgrounds. These findings shed light on complex dynamics of institutional change at technology companies.
\end{abstract}

%%
%% The code below is generated by the tool at http://dl.acm.org/ccs.cfm.
%% Please copy and paste the code instead of the example below.
%%

%%
%% Keywords. The author(s) should pick words that accurately describe
%% the work being presented. Separate the keywords with commas.
\keywords{AI ethics, responsible AI, neo-institutionalism, Science and Technology Studies, organizations, decoupling, institutional change}

%% A "teaser" image appears between the author and affiliation
%% information and the body of the document, and typically spans the
%% page.
%%\begin{teaserfigure}
%%  \includegraphics[width=\textwidth]{sampleteaser}
%%  \caption{Seattle Mariners at Spring Training, 2010.}
%%  \Description{Enjoying the baseball game from the third-base
%%  seats. Ichiro Suzuki preparing to bat.}
%%  \label{fig:teaser}
%% \end{teaserfigure}

%%\received{6 February 2023}
%%\received[revised]{12 March 2009}
%%\received[accepted]{5 June 2009}

%%
%% This command processes the author and affiliation and title
%% information and builds the first part of the formatted document.
\maketitle

\section{Introduction}
Multiple polls have found that public trust in and positive sentiment toward the technology sector has fallen significantly in the US and globally in the past decade \cite{west_techlash_2021, gallup_techlash_2020}. Technology experts have identified the stakes of incidents like Cambridge Analytica \cite{chang_facebook_2018} or algorithmic amplification of disinformation online \cite{menczer_information_2020} as potentially weakening society’s democratic institutions \cite{anderson_many_2020}. By 2018, the Financial Times referred to “techlash” as the best word to summarize the year, claiming that “this year will be remembered as the moment big tech faltered” \cite{foroohar_year_2018}. The wider public’s distrust culminated in multiple instances of Silicon Valley executives being called to testify before Congress; topics ranged from how social media companies had handled misinformation ahead of elections \cite{frenkel_big_2018} to potentially anti-competitive business practices in light of calls for anti-trust action \cite{romm_amazon_2020}.

As the “techlash” movement gained traction, the topic of “ethics” in computing, artificial intelligence, and machine learning took center stage. While what counts as computing ethics is itself vague and contested \cite{metcalf_owning_2019}, here we define it as a set of principles, standards, and processes seeking to prevent harm in the development, implementation, and impact of computing technologies. These include---but are not limited to---values such as fairness, accountability, transparency, privacy, security, and sustainability. In recent years, the field of AI ethics saw massive growth, with the establishment of several conferences and journals dedicated specifically to this topic. Many companies began to “develop ethics ‘in-house’” \cite{phan_economies_2021}, hiring  experts to lead research agendas related to fairness, transparency, and accountability. Yet these initiatives were rapidly criticized as “ethics washing” or “ethics theater,” which Johnson defined as “the practice of fabricating or exaggerating a company’s interest in equitable AI systems that work for everyone” \cite{johnson_how_2019}. The launch and growth of internal corporate research on ethical AI was further challenged as a strategy to “tame” ethics, transform it into a form of capital to protect reputation, and subordinate ethical debates to corporate interests \cite{phan_economies_2021, phan_economies_2022, metcalf_owning_2019}. Others highlighted the challenges associated with addressing ethical issues in AI, including the difficulty of formulating ethical values into quantifiable parameters, the lack of consensus about what is ethical, and the opaque nature of machine learning algorithms \cite{polonski_hard_2018}.

Beyond these heated debates, however, we still know relatively little about the day-to-day experiences of AI ethics workers and the internal processes through which technology companies implement accountability mechanisms and ethical safeguards in the development of artificial intelligence—the ways in which they “walk the walk” instead of simply “talking the talk.” As such, we present one of the first empirical investigations into AI ethics on the ground in the technology sector. To better understand the challenges to the implementation of AI ethics initiatives, we conducted a qualitative study of responsible AI initiatives within technology companies, including 25 interviews with ethics workers and observations in industry workshops and training programs. Drawing on neo-institutional theory, we investigate the dynamics of so-called “decoupling” of AI ethics initiatives in the technology industry: the divergence of publicly articulated formal structures from intended outcomes and on the ground practices \cite{meyer_institutionalized_1977}. We analyze AI ethics workers as institutional entrepreneurs \cite{dimaggio_interest_1988} who, in the absence of structural change due to decoupling, take on the responsibility of organizational change by using persuasive strategies and diplomatic skills. 

Despite predominantly positive reactions to their stated goals within their companies, we find that ethics entrepreneurs faced multiple sources of friction to the implementation of AI ethics. First, they struggled to have their issues prioritized in an institutional environment where power structures were centered around the launches of new software products. As a result, ethics interventions that might have stalled launches or taken up too much time were at risk of going unaddressed. Second, workers were incentivized by metrics associated with financial gain from frequent product innovation, but results associated with ethics interventions were difficult to prove or quantify. Third, we document how constant reorganizations or “reorgs”—a staple of Agile development, a style of product development that emphasizes modular tasks, fast releases, and frequent pivots—made it difficult for ethics entrepreneurs to access institutional knowledge and maintain relationships central to their work. Consequently, we show how individual ethics entrepreneurs ended up taking on significant personal risk when speaking out against potential harms. This was  particularly damaging to workers from marginalized backgrounds, which is paradoxical given the alleged aims of the field of AI ethics. These findings offer a snapshot of the everyday challenges that ethics entrepreneurs faced as they attempted to implement institutional change at technology companies. We discuss implications for the future of computing ethics within organizations.

\section{Literature Review}
\subsection{The Field of AI Ethics}
In the past several years, the field of AI ethics has undergone extraordinary growth both in industry and academia. Companies have responded to external public pressure to act and reform in several ways: they have published principles by which they pledge to abide, hired social science experts to conduct research and compliance for their companies, and employed software engineers to develop technical solutions for these issues. These technical and social science experts have undertaken many public-facing initiatives that correspond with their internal work. Such initiatives include creating technical toolkits to evaluate fairness and to help make machine learning algorithms “explainable,” making these toolkits available for others to use, publishing blog posts detailing steps they have taken to address existing issues, and holding webinars to explain their “responsible AI practices” to others. 

Yet the apparent growth of the field and these public actions have raised questions about their efficacy. Scholars and practitioners criticize AI ethics principles, which articulate the importance of values like fairness, transparency, and accountability in building AI \cite{fjeld_principled_2020}, for several reasons. While some scholars argue that these principles are insufficient on their own to effect change, in part because they are toothless and vague \cite{mittelstadt_principles_2019,munn_uselessness_2022}, others challenge the basis of the principles themselves. For example, Greene, Hoffman, and Stark highlight the underlying assumptions of AI ethics principles as grounded in a limited and technologically deterministic view: many of the statements take for granted that AI and machine learning models will inevitably be built and change society \cite{greene_better_2019}. 

Fewer instances of research have examined the implementation of these principles, tools, and practices on the ground, despite the abundance of work on AI principles and technical proposals in response to AI ethics quandaries. Some exceptions include interview and survey studies that identify the tooling needs of industry practitioners \cite{holstein_improving_2019} and offer policy recommendations to improve the quality and impact of algorithmic audits \cite{costanza-chock_who_2022}. Such studies typically focus on tangible technical and policy recommendations. For example, Holstein et al. identify the need for developer tools explicitly designed to interactively guide data collection, curation, and augmentation \cite{holstein_improving_2019}. Similarly, Costanza-Chock et al. note a widespread lack of “auditee buy in” for algorithmic audits, with 80\% of surveyed auditors reporting that they have made recommendations that were not implemented \cite{costanza-chock_who_2022}. Yet they do not more deeply investigate the organizational dynamics that lead to this outcome.

One notable exception is Metcalf, Moss, and boyd’s analysis of the institutional logics that conflict with AI ethics work: technological solutionism, meritocracy, and market fundamentalism \cite{metcalf_owning_2019}, which are all prevalent in technology companies (see also \cite{avnoon_contextualizing_2023}). Metcalf, Moss, and boyd argue that these institutional logics create “potential for… pitfalls introduced to prevent the work of ethics owners from addressing the broader goal of a more just algorithmic and data-driven world.” Their study identifies the ideological contradictions inherent in AI ethics implementation. Here we build on this work and seek to map out the concrete organizational roadblocks that instantiate and complement these abstract logics when AI ethics employees aim to implement accountability and ethics safeguards. Building on previous calls to study artificial intelligence “in practice” \cite{christin_algorithms_2017}, we take a grounded approach, examining ethics workers as individuals embedded within complex organizational dynamics.

\subsection{Institutional Change: Decoupling and Institutional Entrepreneurship}

To contextualize AI ethics in technology companies, we draw on
neo-institutional theories of organizational change. The neo-
institutional literature focuses on two sources of organizational change: exogenous “jolts” or environmental pressures such as social upheaval, technological disruptions, competition, or regulation \cite{meyer_adapting_1982, greenwood_institutional_2006}; and endogenous action from actors embedded within organizations \cite{dimaggio_interest_1988}. 

According to the neo-institutionalist approach, in response to environmental pressures from external actors, organizations develop rituals and ceremonial formal structures in their attempts to gain legitimacy \cite{dimaggio_iron_1983}. Yet on the ground practices and outcomes often diverge from these formal policies and procedures, as “activities are performed beyond the purview of managers” and “goals are made ambiguous and vacuous” \cite{meyer_institutionalized_1977}. For example, school principals may introduce a formal structure to improve education at their schools, but they are not generally present in the classroom where teachers work to ensure that teaching practices follow formal policies. If practices are implemented, measures of achievement may be deemphasized to obscure whether the practices produce the intended outcome. This phenomenon in which policies, practices, and outcomes are not necessarily aligned is known as “decoupling” or “loose coupling” \cite{meyer_institutionalized_1977}. Decoupling can take several forms. It may look like a gap between policy and practice, where the formal policy has little or no relation to the daily practices; or it may involve a disconnect between means and ends, where the formal policy is implemented in practice but is only loosely tied to “outcomes” \cite{bromley_smoke_2012}. 

Given that decoupling undermines the efficacy of external pressures in stimulating actual reform, organizational change is more likely determined instead by individuals, or “the networks of social behavior and relationships which compose … a given organization” \cite{meyer_institutionalized_1977}. According to the concept of institutional entrepreneurship, “new institutions arise when organized actors with sufficient resources see in them an opportunity to realize interests that they value highly” \cite{dimaggio_interest_1988}. Institutional entrepreneurs must first break with the dominant institutional rules, practices, and logics, and then institutionalize the alternative they are championing. These processes are political negotiations that require institutional entrepreneurs to have the “capacity to imagine alternative possibilities.” They also require that institutional entrepreneurs have the skills to strategically frame their projects to “mobilize wide ranging coalitions … and to generate the collective action necessary to secure support for and acceptance of institutional change” \cite{garud_institutional_2007}. By mobilizing different coalitions and generating collective action internally with their colleagues, institutional entrepreneurs may be able to “recouple” policies, practices, and outcomes, thereby effecting organizational change \cite{espeland_struggle_1998}.

How exactly do institutional entrepreneurs mobilize coalitions, generate collective action, and institutionalize change? Greenwood, Suddaby, and Hinings expound on the process of “theorization” as an essential stage of institutional change \cite{greenwood_theorizing_2002}. Theorization involves two tasks: the specification of an “organizational failing,” and the justification for a proposed solution or change. This process simplifies the properties of new practices, making them available for broader diffusion if “new ideas are compellingly presented as more appropriate than existing practices” \cite{greenwood_theorizing_2002}. We use the concepts of “decoupling” and “institutional entrepreneurship” to understand AI ethics workers who attempt to recouple ethics interventions through theorization, rather than accept employment in symbolic structures as “ceremonial props” \cite{vo_microfoundations_2016}.

\subsection{Decoupling, Entrepreneurial Labor, and Ethics in Silicon Valley}

As noted above, technology companies publicly disclose  policies (in the form of AI principles) and tools (in the form of toolkits, audits, and documentation). However, it remains unclear whether these initiatives are actually implemented and tied to concrete outcomes.

This in turn is part of a broader development in modern capitalism. As Bromley and Powell explain, instances of means-ends decoupling have become more common in contemporary organizations: increased societal pressure toward transparency and accountability make policy-practice decoupling likely to be perceived as a moral failure, as opposed to furthering legitimacy. They suggest that means-ends decoupling is more prevalent in contexts when outcomes are hard to measure, “when internal constituents champion an external cause,” and when the rationalized environment is fragmented (perhaps due to societal pressure that results from visibility and the perception of public interest) \cite{bromley_smoke_2012}. Drawing on Bromley and Powell’s characterization, the case of AI ethics is a good candidate for means-ends decoupling.

Furthermore, previous literature has shown that globalization and the rise of the computing industry produced more decentralized and less hierarchical organizational forms, which thrived in Silicon Valley \cite{powell_capitalist_2001, saxenian_new_2006}. As Neff and Stark write, “when the challenges of responsiveness are too great for institutional routinization, organizations’ bureaucratic structures are destabilized. Heterarchies— flatter organizational structures with distributed accountability, decentralized decision making, and multiple, often competing, goals— emerge" \cite{neff_permanently_2004}. In this flexible, flatter environment, everyday organizational decision-making takes a different shape: consensus is developed through dialogue and influence rather than one’s formal position in the hierarchy \cite{heckscher_defining_2011}. Informality and flexibility dominate in service of a high-change environment that allows for quicker product development, as opposed to formality, rules, and strict boundaries which may slow production. The relationship between this project-based organizational structure and the necessity of entrepreneurial labor in Silicon Valley is demonstrated by previous research, detailing the work of internal advocates who influence and persuade others in order to ensure values-oriented change \cite{wong_UX_2021} as well as address privacy \cite{tahaei_privacy_2021} and security issues \cite{tondel_security_2020, haney_cybersecurity_2018}.

Given the relevance of decoupling to the case of AI ethics and the flatter organizational structure characteristic of Silicon Valley corporations, we need a better understanding of the interpersonal dynamics upon which institutional change in the technology industry depends. As our study documents, institutional entrepreneurship is a relevant lens by which to analyze ethics workers: we analyze them as \textit{ethics entrepreneurs} who, lacking organizational support, must negotiate with coworkers to bring about institutional change in the field of computing.

\section{Methods}

To examine the implementation of AI ethics in the technology sector, we proceeded in several steps. First, we familiarized ourselves with the field of AI ethics through a review of the literature and through online and in-person observation of various public talks, conferences, and webinars related to AI ethics (many of which we learned about through word of mouth and publicization on Twitter and LinkedIn). We spent about 40 hours at these sites in 2022. Next, we engaged in a mapping of the field and compiled a spreadsheet of relevant actors to interview. We added to the list iteratively to maximize variation in terms of position, sociodemographic characteristics, and type of company, culminating in a total of 85 professionals. We reached out systematically to people for interviews. We then reached out to other potential participants via snowball sampling (asking participants for referrals). We also relied on the LinkedIn recommendation algorithm to identify other respondents \cite{christin_ethnographer_2020}.

In total, we conducted 25 interviews with employees, academics, and consultants, primarily ethics workers currently or formerly employed as part of responsible AI initiatives at technology companies. Four of the 25 interviews were conducted with people who worked in machine learning without a focus on responsible AI to gain perspective from people with whom ethics workers were interacting with. Many people we contacted declined to participate due to non-disclosure agreements (NDAs), advisement from the legal team at their company about intellectual property, other concerns around privacy, or undisclosed reasons. All the interviews were conducted by a single researcher, the lead author. In total, we contacted over 150 professionals via email and social media direct messages.

Our sample was highly educated, as 24 of the 25 participants had obtained graduate degrees. Their expertise ranged from technical software engineering and computer science to humanities and social science backgrounds. The roles spanned technical and non-technical positions, including titles like “Researcher,” “Engineer,” and “Program Manager.”  The teams they were a part of varied as some companies had dedicated responsible AI teams, while others belonged to Trust and Safety or civic integrity teams. The companies they worked for included consulting and technology firms (ranging from about 6,000 employees to hundreds of thousands of employees). 60\% of our sample identified as men. About half identified as white, and other interviewees were Asian / Pacific Islander, South Asian, Middle Eastern, and Black. Most interviewees were in their late 20s to mid-40s. 

Interviews were semi-structured and lasted between 30 to 70 minutes over video-conferencing platforms. We told participants we were interested in the organizational barriers and challenges of implementing AI ethics initiatives and interventions in industry. In accordance with our IRB-approved consent procedure, we asked for oral consent to be interviewed and recorded. We recorded 21 of the interviews; four participants declined to be recorded due to the sensitive nature of the interviews and the perceived risk to their position, in which case detailed notes were taken. The recorded interviews were transcribed, and the names of the interviewees and companies at which they held a position were changed to preserve anonymity. 

To gain insights into our broader question about organizational barriers and dynamics, we relied on abductive analysis \cite{tavory_abductive_2014}. Our original questions were relatively open, focusing on participants’ experiences with regards to tools, interpersonal dynamics, and organizational structure. We analyzed this data inductively to produce insights about structural problems at the level of the organizations that the participants worked in. We then revisited the data with our neo-institutionalist framework and questions in mind. In addition to sharing personal experiences, many of the highly educated participants tended to abstract out beyond their observations to a higher-level analysis of the field as well, and we took note of the distinction between these types of data in our analysis.

\section{Findings}
\subsection{Becoming An Ethics Entrepreneur}

Overall, ethics workers described the product teams that they worked with, including engineers and product managers as, by and large “agreeable.” Kate, a PhD student who had worked at multiple companies in AI ethics-related roles, described people’s reactions: “Everyone’s like ‘Oh, I want to do this… Like, I’m happy to help you. I’m happy to participate.’” Others were pleasantly surprised by the positive reception to their efforts, with one participant highlighting the increased public awareness of the negative impacts of AI as compared to five years ago. Despite these positive reactions, ethics workers still faced challenges when it came to implementation of their recommendations. A lack of buy-in from leadership meant that ethics workers lacked organizational support, such as authority and resources, to fulfill their roles, perhaps a reflection of the intended symbolic nature of the role. To accomplish their goals, ethics workers became ethics entrepreneurs, relying on interpersonal skills to build active support. 

Many participants mentioned the main problem they faced, which came with several downstream effects: a lack of support from leadership. Sienna, a social scientist who had worked at two large technology companies, pointed out that a challenge at big organizations was getting “buy in,” which was especially difficult given the uncertainty surrounding legal frameworks and the internal specifications of what “fairness” might even mean in machine learning. As she explained:

\begin{quote}
The last challenge at big orgs is getting company-wide buy-in—that means leadership support and bottoms up engineering support for this work. Sometimes it’s hard to do that because of company priorities, and there’s always resource constraints. There’s not clarity in the law, and there’s not clarity in research yet, so it’s difficult to make the case that we need to do XYZ… It’s hard to get buy-in when you don’t have that clarity. 
\end{quote}

Morgan, a former engineer and ethics worker at a Big Technology company, echoed these feelings: “Sometimes, no VP is willing to advocate for our team getting in the way of somebody’s product launch because there’s some ethical issue… There’s no incentive for people at the top to like help the people who do have ethics experience and background make waves in the company and change things.”

We observed this lack of buy-in firsthand through an interview with Siva, a Director-level employee at a Big Tech company. Siva was not working in a role directly related to AI ethics or responsible AI; we spoke to him to gain a better contextual understanding of how ethics entrepreneurs might be perceived. He emphasized that almost everything he oversees requires automation. When asked if he had ever consulted the fairness team at his company, he bristled at the suggestion: “I haven’t had the need to consult the fairness team. No, it’s just not how [the company] works. It’s not possible for small teams to keep on auditing the entirety of products. Each team has its own standards and procedures, launches are reviewed, there are launch metrics.” When asked if he had ever noticed a potential issue related to ethics or fairness close to a launch or after an AI product had already launched, he flatly denied any such occurrence. In response, he said the company was “well-managed” and so “issues like that rarely surface.” Notably, the company and vertical Siva was employed in, which we do not disclose to preserve anonymity, has been publicly critiqued multiple times by academics for racial bias in its algorithms.

Given the difficulties around securing buy-in from leadership, it followed that there was often insufficient organizational support to ensure that AI products were developed responsibly. Ayesha, a Black woman PhD student who had worked at multiple technology companies, gave the caveat, “To a certain extent, it’s [leadership buy-in] there; otherwise, the team wouldn’t have formed.” However, for the team to be truly effective, they needed more leadership buy-in in order to ensure that resources (such as time and personnel for data labeling) were dedicated to these efforts. They also needed leadership to formalize practices as requirements (such as reviews or audits early in the development cycle) and to grant ethics workers authority to mandate changes to products before launches.

Without authority granted to ethics employees, formalization of an ethics review process at an early stage in the development cycle, or resources specifically dedicated to responsible AI, the onus for change falls onto individuals who must use various strategies to influence others. Following the framework of institutional entrepreneurship, we thus analyze AI ethics workers as ethics entrepreneurs. This framework helps us reconcile the role that ethics workers play in effecting change within the organization, the dominant institutional logics that they must fight against, and the inconsistencies in whether ethics interventions are actually applied. While critics argue that ethics is subordinated to capitalist interests \cite{metcalf_owning_2019, phan_economies_2021}, we examine how ethics interventions are nevertheless implemented, albeit inconsistently, as a result of the efforts of ethics entrepreneurs. The ethics entrepreneurs we spoke to relied on their individual relationships across the company, advocates within middle management, and skills involving persuasion and diplomacy to be effective at their jobs. The institutional entrepreneurship framework demonstrates the dependence of ethics work on individuals, in contrast to the corporate façade of structural reform but also in opposition to the overly simplistic interpretation of ethics washing. 

Within this institutional environment, ethics entrepreneurs could intervene successfully using interpersonal strategies such as identifying and motivating allies or reframing problems in terms of mutual interest. Yet we find that issues could just as easily go unaddressed depending on personal interests, interpersonal dynamics, internal politics, and more. For instance, ethics entrepreneurs often described product team members reacting defensively when concerns were raised, others passive aggressively acknowledging critiques but stating that they would not be acting on them, and some perceiving the work as “somebody else’s problem” and trying to offload all work to ethics entrepreneurs (which was problematic when the product team held the expertise on the machine learning model). One ethics entrepreneur working in Trust \& Safety, Dave, explained how he explicitly networked within teams to find out, “Is there a certain personality on the team that just tends to have a lot of influence?” Dave would work with the manager of the team and an ML engineer who knew the team to figure out who needed to be in the room when he presented results. Ayesha similarly described the delicacy with which she had to approach others in order to do this work:

\begin{quote}
It’s hard to ask them [to do things] because of politics. You want to have good relationships with engineers. You’re supposed to be in a role where you’re not taking too much of their time, and it’s supposed to be in partnership. You’re supposed to ping them, but you can’t ping them too much. There’s a lot of diplomacy that has to happen.
\end{quote}

Overall, with the responsibility to make change dependent on individuals and small-scale interactions, participants described a wide variation in their experiences. Some described being met with skepticism on the one hand but also “deeply curious and empathetic people” on the other. Given these differences between interlocutors, ethics entrepreneurs were more likely to engage deeply with people on product teams who cared about the work and willingly participated while sometimes disengaging entirely with teams who were indifferent to or frustrated by these efforts. Overall, the decoupling of policy, practice, and outcomes as well as the lack of organizational infrastructure meant that interpersonal dynamics could take precedent in how a matter unfolded. This resulted in inconsistency in how and whether specific issues were addressed. 

\subsection{Power Lies with Product}

As we saw, means-ends decoupling can emerge when there is a fragmented rationalized environment: the firm operates according to one rationalized logic but is also responding to environmental pressures that operate according to a different rationalized logic. As such, any reforms responding to environmental pressures are unrelated to the core tasks of the organization and thus siloed \cite{bromley_smoke_2012}. In the case of AI ethics, the technology industry assumes product innovation as the core task of the organization, while external pressures emphasize values like fairness and justice. In this section, we present the ways in which power lies with product and, regardless of whether or not ethics is siloed per se, the goal of product innovation supersedes goals related to responsible AI. The authority to make product decisions thus comes from product managers. In response to the supremacy of “product” within the organization, ethics entrepreneurs attempted to negotiate with product managers for resources, theorize ethics-oriented interventions as functionally superior or “pragmatically legitimate” \cite{greenwood_theorizing_2002} by linking them to product quality, and advocate for early integration into the product development cycle.

Participants noted that power was often located with product managers because, as one participant said, “they often bring all of the right resources together to drive product development.”  John, who interacted with compliance-related employees, described the kinds of resource constraints ethics entrepreneurs would face: “Are there funds available for teams to collect a new dataset if the one that they used was problematic in some ways? Or like funding for like re-labeling, or re-annotating an existing dataset? Or even just like having the time and personnel to conduct a fairness evaluation?” If resources were not allocated for ethics work, ethics entrepreneurs would sometimes attempt to convince the product team to allocate funds out of their project budgets for these kinds of activities.

As another participant noted, product managers “are ultimately responsible for making sure there’s time available.” Accordingly, many ethics entrepreneurs highlighted that they would encounter resistance from them if and when their recommendations were likely to delay timelines for product launches. Sam, who previously worked at a Big Tech company but transitioned to a non-profit, said while she perceived teams as amenable to her suggestions, “They might be like, ‘We don’t have time for this, we have to get to launch.’… No one was antagonistic to it, there was more just this pressure to get this thing out.” Specifically, leadership would set timelines for product launches in order to beat “investor expectations” through frequent product innovation. The focus on launch deadlines as well as the impetus to “move fast” and have a “bias for action” (corporate values at Meta and Amazon respectively) sets expectations within the company that rapid product innovation is the priority above all else.

Multiple participants posited that efforts would be more successful if leadership were to mandate integration of ethics initiatives into early stages of the product development cycle. Respondents noted that they had positive experiences when engaging in a consultation or review of a product early in its design, such as during the ideation or conception phase. In those cases, suggestions could be incorporated early on and could shape the direction of the product. However, teams often consulted responsible AI team members at the end of a product development cycle, for example after a model was already trained but before it was deployed. This would result in scrambles to course correct just before a launch deadline if it was even feasible to do so. Gabriel, an ethics entrepreneur at a large technology company with a background in philosophy, noted,

\begin{quote}
    The thing that a team might be annoyed about is the fact that the work is delayed, or like major suggestions are being made very late on in the process. But I think like most of that anger could be taken away through a very early engagement. So if… these points had come up, and the direction of the project had changed, there probably wouldn't be such a tension…. But the question to ask there is like, why is it the case that the team ended up going to ethics review so late in the game?

\end{quote}

Perhaps Gabriel’s question could be answered by our application of neo-institutional theory: that ethics and product follow different and separate rationalized logics that silo and buffer activities from each other.

How then do ethics entrepreneurs overcome challenges associated with the supremacy of product? Some ethics entrepreneurs take a pragmatic route, focusing on making automated ethics tools as easy as possible to use in order to reduce “friction” and achieve higher adoption rates amongst product teams. Others were able to influence others by reframing any problem as an issue of product quality. For example, Kate said,

\begin{quote}
    I think the vernacular we use around these topics is really important, especially when working with machine learning practitioners… I’ll try to pick up as much of the internal culture and internal jargon as I can. I very rarely say 'fairness' or 'ethics' or anything like that because like, that’s a loaded word…  I really try to go to product teams with the idea that, ‘I'm here to help you make a better product.’ And in my experience, the way organizations work, everyone can get behind making better products.

\end{quote}

Using product quality as a justification demonstrates one way for ethics entrepreneurs to integrate ethics activities more seamlessly into the core activities of the organization. It also confers legitimacy to a new practice that developers may not see the value of.

Finally, while ethics entrepreneurs attempt to integrate ethics into the organization’s core activities, others expressed a more critical view of the structural conditions surrounding frequent product innovation. One informant who had done AI ethics-related consulting for different technology corporations critiqued this focus on constant innovation as part of a systemic issue with technology and venture capital: 

\begin{quote}
    We see all of these products that are based on some sort of speculative reward. And these products have all of this runway based on the amount of investment in them. And so, they’re able to go out there into the real world and produce harm without necessarily having a real market need. Like nobody needs better ads…. But if they’re gonna get \$500 million of VC funding to give better ads, they’re going to build something that promises to give better ads, whether it does or not.

\end{quote}

Consistent with this critique of needless and constant product innovation, multiple participants raised an example of an ethics issue where the solution was not to correct the model, but rather simply not to use AI for the feature at all. 

\subsection{The Problem with Performance Metrics}

We have seen how product innovation was perceived as the top priority for many technology organizations, superseding ethics concerns; one way in which the prioritization of “product” is operationalized is through a company’s metrics. 

Ethics entrepreneurs often referred to “North Star” metrics and team sub-metrics, typically issued by analytics software companies such as Mixpanel and others, as central to the negotiations around their work. Some product teams held the assumption that any intervention to improve fairness, for example, was going to “degrade” their models, make them less accurate, or affect their ability to meet revenue targets. In those cases, ethics entrepreneurs had to spend time “pre-emptively assuaging their fears” and make sure their suggestions did not “make things harder for that team.” Dave, who worked in Trust and Safety, said he would frequently encounter the concern over less accurate models but did not find that AI ethics suggestions actually resulted in degraded models in practice. 

Many ethics entrepreneurs further described a pressure to provide quantitative evidence for their own ethics or responsible AI-related goals. Several said that they were currently focused on defining metrics for fairness or ethics, because that was the existing structure through which work was incentivized. Dave explained:

\begin{quote}
    No matter how much people want to do a good job, at the end of the day, quarterly goals will get in the way. Whether it’s revenue or monthly users that your job depends on. Early advice I got was, if you could ever get the CEO to say, 'Ok one of our North Star metrics is related to ethics/fairness,' that would have just changed everything.

\end{quote}

At the same time, multiple ethics entrepreneurs described their focus as working to determine “ethics” or “fairness” metrics in response to the question, “What is ethics and how well are you doing it?” However, many characterized this as an inherently difficult problem. While pursuing his PhD, James interned at a Big Tech company whose business model was centered around engagement. He depicted his own role as measuring the merits of introducing friction to make their AI products more responsible:

\begin{quote}
    Our job was to find such concrete evidence of connections between the platform and bad things happening in real life… that we could convincingly say, ‘You will lose more money… if you don’t throttle, than you will lose by throttling.' And it’s really hard to have evidence that concrete, like predicting the future. And so in my time, I don’t think I witnessed any while I was working there.

\end{quote}

In this example, the North Star metrics around engagement were so highly prioritized that any recommendations to reduce those metrics required irrefutable quantitative evidence. At the same time, uncovering evidence of this kind necessitated data work that was not well supported by the existing data infrastructures at the company. As such, without this evidence, the default was to continue to prioritize product innovation that increased engagement. Furthermore, much as Bromley and Powell had warned \cite{bromley_smoke_2012}, ethics entrepreneurs were directed to spend time and resources measuring outcomes for a goal that might not be measurable, as opposed to spending that time implementing practices that were likely to further responsible AI goals.

Relatedly, when ethics entrepreneurs introduced practitioners to fairness toolkits that would give a better sense of how the model performed for different demographic groups, they would react by asking for a specific target they needed to meet. One participant recounted his experience recommending the use of fairness toolkits to engineers who would respond by asking him, “What's the threshold? Is it 80\%? 85\%? You know, what should I do?” These questions often did not have clear answers or legal standards and could require more philosophical inquiry at the foundation of what it meant for the model to be fair and ethical.

\subsection{The Problem with "Reorgs"}

Multiple participants mentioned the frequency of reorganizations, or “reorgs,” which rearranged the structure of their teams and/or the teams they worked with. Some ethics entrepreneurs described it as a source of stress for product teams as well. Theoretically, reorganizations would support “Agile” product development, a popular management style for software engineering. As Posner explains, Agile was “designed for ultimate flexibility and speed, it requires developers to break every task down into the smallest possible unit. The emphasis is on getting releases out fast and taking frequent stock, changing directions as necessary” \cite{posner_agile_2022}.  Frequent reorganizations in this environment would indicate that the organizational structure is responsive to the shifting needs of the company and projects. James, the PhD student intern at a Big Tech company, sourced this philosophy of constant reorganizations as part of engineering and startup culture:

\begin{quote}
    The reshuffling and the reordering is supposedly, you know, companywide policy for continuous optimization and continuous iteration…. That top-down tech culture leads people to think that you could do constant A/B tests on your personnel and your corporate structure, just like you do constant A/B tests on your platform for user engagement checks, and just see what fits, what works better. And like every time you see an adjustment that you think might improve metrics just a little bit, then you change just a little bit.

\end{quote}

James explained that his team had three different bosses over the course of his three-month internship: he was unable to access institutional knowledge that might have helped him do his job better and avoid repeating previously attempted projects. Importantly, in the context of our other findings, it could also mean that relationships that ethics entrepreneurs had taken time to build and depended on for their work were disrupted, creating additional friction and inefficiency to implement ethics.

\subsection{The Individualization of Risk in AI Ethics}

Our study identifies several characteristics of technology companies that lead to ethics work being inconsistently taken into consideration by individuals. We find that ethical questions are de-prioritized in product development cycles, disincentivized by metrics, and disrupted by constant reorganizations. These challenges come with important ramifications in terms of inequality and individual risk for workers invested in fairness, equity, and ethics. 

First, given that North Star metrics aligned teams within the company toward a common goal, individuals were likely to be evaluated on their contributions to those metrics rather than ethics-related objectives. As such, individuals’ promotions and careers are at stake as they try to meet their launch deadlines and improve metrics, as opposed to doing due diligence on ethics concerns that might stall a product. While evaluation and promotion based on metrics was one area of concern and could result in AI Ethics work being disincentivized, many ethics entrepreneurs described risks they felt more broadly as well. Some of these perceived risks were described above, but even ethics entrepreneurs who did agree to participate in our study voiced various hesitations about participating, being audio recorded, being directly quoted, and otherwise being identified by their interviews in this publication, especially if they were still employed by technology companies in this field. Furthermore, participants described the acute personal risks and pressures they felt as they went about their day-to-day jobs. Sam commented, “Being very loud about putting more brakes on it was a risky thing to do. It was not built into the process” and noted that it required appealing to leadership. Likewise, Dave, described why he left his previous company, where he had started their AI ethics initiatives:

\begin{quote}
    I was going to be in a position where if something did [an issue surfaced close to or after launch], I would be the one to see it and basically need to be the whistleblower. And I realized there were really no protections in place for me. And the stress level that was causing was not really being acknowledged… [Ethics workers] can be in a position of being put under unreasonable pressure to basically need to decide or at least be the one to deliver the bad news to teams. Like, ‘Yeah, you should pull the plug on this before you launch it.’ … It’s just not acknowledging how much pressure that puts on people who are often not senior enough to be making those decisions.

\end{quote}

While Dave felt this pressure as a white man, many participants highlighted that marginalized people were often more likely to experience these adverse kinds of pressure. Ayesha, a woman of color, observed that “the people who are actually heavily invested in it as like---as a call to action, as a vocation, and as a means for … social change, are the ones who inhabit marginalized identities,” noting that experiencing discrimination themselves means that the work “hits differently.” She noted that she often felt she had to speak up “whenever like I thought that something terrible was happening, not just with the team, but the company at large.” Many participants also referenced Timnit Gebru, a Black woman computer scientist who was employed at Google and whose departure from the company made headlines in the media, as evidence of the risk that people from marginalized backgrounds take on in this field \cite{wong_more_2020}.

The male-dominated, engineering-centric world of Silicon Valley may have made speaking on fairness and ethics related issues as someone from a marginalized background particularly difficult. Ayesha described some antagonism as “part of an engineering norm that says it’s okay to tear people down instead of providing thoughtful criticism.” Some of the pushback she had received in professional spaces reminded her of “reddit threads that are really… white supremacist and incel-y.” She felt that because of the position she was in, she “couldn’t just shut it down immediately” and instead had to be “very thoughtful and diplomatic and graceful about it.”

To summarize, the way ethics and responsibility principles are incorporated at many technology corporations puts the onus on individuals to take on work as opposed to relying on organizational processes which may produce more consistent outcomes. As a result, it puts stress and pressure on individuals who advocate at great personal risk to their own careers. This can be particularly damaging to people from marginalized backgrounds, who bring an underrepresented perspective and whose lived experiences with discrimination may compel them to speak up. Not only does this further harm already marginalized people by adding additional burdens and barriers to their success at work, it also has the potential to push them out of the organization entirely from sheer exhaustion or as the result of retaliation.

\section{Discussion}

\subsection{The Contradictions of Formalization}

In our study, the description of ethics work that emerged was one of an informal process, primarily undertaken by the individuals we analyze as “ethics entrepreneurs.” That said, a few participants mentioned that their companies had begun to implement more formal measures to make responsible AI more process-driven through mandatory impact assessments at the start of the product development cycle. If the product was determined to be a “sensitive use” through the impact assessment, this would then trigger review and approval by a committee. This process was described as “checks and balances” that would “catch things that might have otherwise gone through the net, especially at this stage where like, not everyone’s upskilled in this space.” Participants usually viewed these favorably, but it was a small minority that described the ethics review process in this way. Most alluded to an absence of formalization or contrasted processes around AI ethics with the case of privacy and security checks. With a formalized process, the bare minimum standard for engagement is raised, and checks on unbridled innovation are introduced. 

Our interviewees voiced mixed feelings towards the formalization and routinization of AI ethics through checklists. They explained that a checklist itself would not necessarily result in immediate change or correction; however, they acknowledged that it would trigger the attention of expert reviewers who could engage product teams in a deeper conversation as needed. While the merits of formalization may be debated theoretically, we note that decision makers could look to older computing-related subfields such as security or privacy to evaluate whether formalization of those frameworks has provided a net benefit to society’s relationship with information technologies: many of our participants looked to these subfields as a positive example to follow. Still, previous literature has shown how both subfields rely on the labor of individual workers within software teams who champion these causes \cite{tondel_security_2020, haney_cybersecurity_2018, tahaei_privacy_2021}.

Another pathway towards formalization that was regularly touted and explored by our participants was the role of regulation in improving AI ethics initiatives. There was a consensus that regulation would undoubtedly incentivize leadership at technology corporations to commit to taking these matters seriously, especially if there were financial penalties for non-compliance. However, there were also concerns that regulations could be too “one-size-fits-all” in ways that could make matters worse, such as requiring all platforms to collect demographic information about their users. 
 
Furthermore, participants expressed that having legal standards could provide not just adversarial pressure on leadership to invest in ethics, but also clarity and guidance for ethics entrepreneurs and product teams. For example, our participant Khan, who worked for many years at a large technology company, said that many areas in machine-learning fairness require more definitional work on “what it means for something to be socially beneficial.” Ethics entrepreneurs and the product teams they worked with wrestled with difficult questions and tradeoffs around fairness. For example, what would be an equitable representation of different demographic groups in algorithmic curation and promotion? Is there an acceptable performance drop in accuracy for different demographic groups if a machine learning model would be beneficial when deployed, but disproportionately beneficial to certain groups? Multiple participants commented that regulations could provide valuable direction and standards to work towards, as opposed to workers internally—and often individually—trying to make difficult and important decisions affecting people at a large scale.

\subsection{Decoupling and Institutional Entrepreneurs}

The neo-institutional literature offered relevant insights into the future of the field of AI ethics and its efficacy. Here, we discuss some potential consequences of decoupling and importance of ethics entrepreneurs in effecting change.

As Bromley and Powell show, means-ends decoupling, for which AI ethics is a prime candidate, has some likely downstream effects \cite{bromley_smoke_2012}. One consequence is that the structure of the organization can become more complex, with specific activities (corresponding to external logics like fairness) compartmentalized and isolated from the rest of the organization (which remains focused on its own core goals and logics). We have seen this play out in our study, with AI ethics concerns deprioritized relative to product goals and metrics. However, in addition to the ethics entrepreneurs we interviewed, we also reached out to ten product managers in our personal networks at Big Tech firms. Of the six who replied to our inquiries, only one had heard of a responsible AI or AI ethics team at their companies. This suggests that ethics entrepreneurs may be more siloed than we have thus far described, though we leave this to be investigated by future research. Another consequence of means-ends decoupling is perpetual reform---either periods of calm interrupted repeatedly by waves of reform, or incremental gradual change---as opposed to massive, stable, system-wide change.

Finally, despite the obstacles we have described, our framing of ethics workers as ethics entrepreneurs does offer some hope for institutional change, in contrast to previous work in the field of AI ethics. Those concerned with the implementation of AI ethics can advocate for more institutional support and whistleblower protections for ethics entrepreneurs. Furthermore, other studies of institutional entrepreneurs indicate that they may be able to gradually institutionalize external logics around justice in order to achieve organizational change. The growth of tech worker organizing is an indication that this avenue may yield results \cite{kramer_tech_2021}. At the same time, we caution against relying on individuals rather than collective action and structural reform to effect change, especially given the incredible risks for workers from marginalized backgrounds.

\section{Conclusion}

This article captures the difficulties of institutionalizing change within technology organizations. Based on a qualitative study of AI ethics workers, we document an environment where policies, practices, and outcomes are decoupled—one where companies “talk the talk” of AI ethics without fully “walking the walk” of their implementation. We show that decoupling and insufficient leadership buy-in can compromise an institution’s ability to address AI ethics issues adequately and consistently. Ethics workers must rely on influence and charismatic authority rather than formal structure in an environment that prioritizes product and its associated performance metrics. As a result, individuals are on their own when it comes to implementing ethics. They must behave as ethics entrepreneurs, pushing a political agenda throughout the organization without meaningful infrastructural support, which often puts them in a position of high personal risk. We show that these risks may be particularly acute for those from marginalized backgrounds, which is ironic given the alleged aims of the field of AI ethics. While individual ethics entrepreneurs may act as catalysts for organizational change, we conclude that the meaningful implementation of ethical concerns within technology companies must grapple with questions of power and inertia in organizational contexts.

\begin{acks}
We thank Stanford Human-Centered Artificial Intelligence Institute (HAI) for their seed grant which funded this work. We also thank Fred Turner, Jay Hamilton, Shira Zilberstein, the Stanford Communication Works in Progress Group, the Stanford Center for Work, Technology, and Organization, and the FAccT reviewers for their helpful feedback. Finally, we thank our participants who trusted us to convey their experiences safely and meaningfully.
\end{acks}

\bibliographystyle{ACM-Reference-Format}
\bibliography{walkthewalk}

\end{document}